\documentclass[prb,aps,twocolumn,showkeys]{revtex4}
\usepackage{graphicx}
\usepackage{amssymb,amsfonts,amsmath}

\begin{document}
\title{Critical exponents and scaling invariance in the absence of a critical point}
\author{N. Saratz$^1$}
\author{D. A. Zanin$^1$}
\author{U. Ramsperger$^1$}
\author{S. A. Cannas$^2$}
\author{D. Pescia$^1$}
\author{A. Vindigni$^1$}
\email{vindigni@phys.ethz.ch}
\affiliation{$^1$Laboratorium f\"ur Festk\"orperphysik, Eidgen\"ossische Technische Hochschule Z\"urich, CH-8093 Z\"urich,
Switzerland}
\affiliation{$^2$Facultad de Matem\'atica, Astronom\'{\i}a y F\'{\i}sica (IFEG-CONICET), Universidad Nacional de C\'ordoba, Ciudad
Universitaria, 5000 C\'ordoba, Argentina}
\date{\today}


\begin{abstract}
The paramagnetic-to-ferromagnetic phase transition is believed to proceed through a critical point, at which power laws and scaling invariance, associated with the existence of one diverging characteristic length scale -- the so called correlation length -- appear. 
We indeed observe power laws and scaling behavior over extraordinarily many decades of the suitable scaling variables at the paramagnetic-to-ferromagnetic phase transition in ultrathin Fe films. However, we find that, when the putative 
critical point is approached, the singular behavior of thermodynamic quantities 
transforms into an analytic one: the critical point does not exist, it is replaced by a 
more complex phase involving domains of opposite magnetization, below as well as $above$ the putative critical temperature.  
All essential experimental results are reproduced by Monte-Carlo simulations in which, alongside the familiar exchange coupling, the competing dipole-dipole interaction is taken into account. Our results imply that a scaling behavior of macroscopic thermodynamic quantities is not necessarily a signature for an underlying second-order phase transition and that the paramagnetic-to-ferromagnetic phase transition proceeds, very likely, in the presence of at least two long spatial scales: the correlation length and the size of magnetic domains.       
\end{abstract}
\maketitle
\onecolumngrid
\begin{figure*}
\includegraphics*[width=1.\textwidth,angle=0]{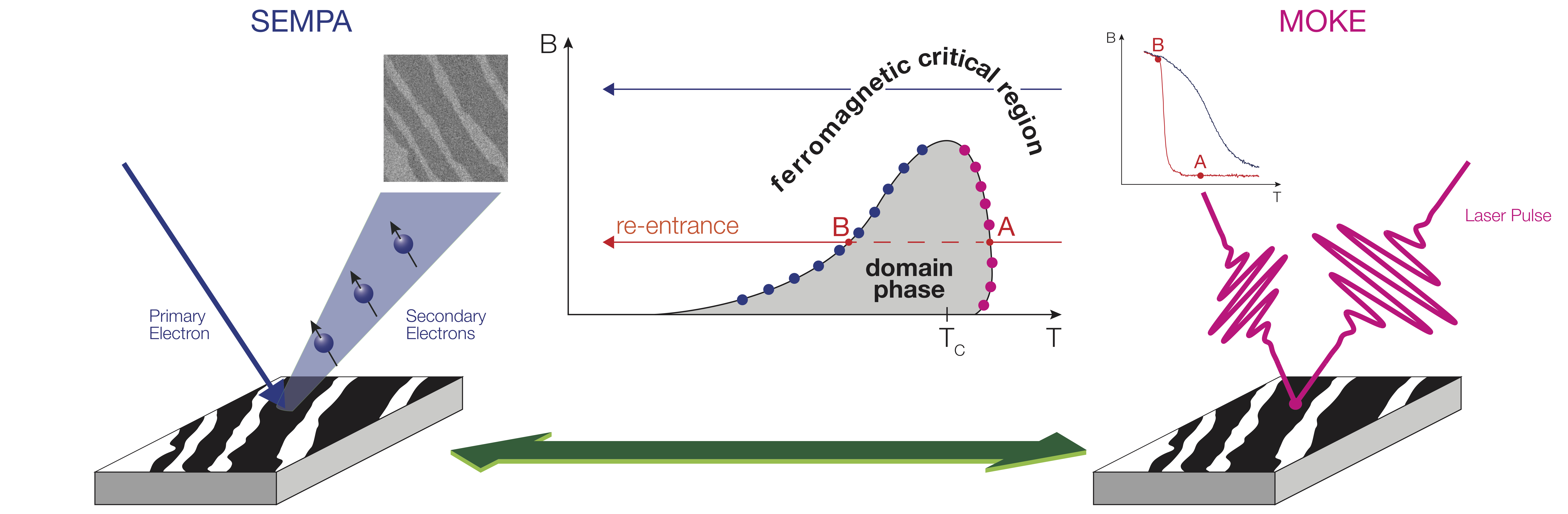}
 \caption{
Sketch representing the complementary use of MOKE and SEMPA techniques to determine the phase diagram of Fe/Cu(001) films. SEMPA allows defining the phase boundaries by direct	visualization of the domain pattern realized for a given pair of values of $T$ and $H$ (e.g., blue dots in the pictorial phase diagram); this approach can, however, be applied as long as the time needed to scan a whole image ($\sim$  minutes) is faster than the typical timescale over which domains move. As the putative $T_C$ is approached, domains become mobile and the contrast is lost. The information extracted from SEMPA imaging was complemented with MOKE measurements of the \textit{global} magnetization, in which a lapse of time in the range of seconds was used to sweep a full magnetization curve at constant $T$. Fuchsia dots exemplify a  boundary between the uniform phase and the domain-patterned phase that can be defined through the analysis of MOKE data described in this work. The blue and the orange paths indicated in the phase diagram produce the two types of magnetization curves reported in the inset on the right and described in section III.A.}
\end{figure*}
\twocolumngrid 
\section{Introduction}  
\vspace{-0.4cm}  
\noindent Our understanding of the ferromagnetic phase transition foresees the appearance of a finite, spatially uniform spontaneous (i.e. occurring in zero applied magnetic field) magnetization as the temperature is lowered below a well-defined critical (Curie) temperature~\cite{Landau}. Strictly speaking, however, a finite three-dimensional body 
(i.e. a body extending over sufficiently large but {\it finite} lengths in all three spatial directions) lacks a finite uniform magnetization at any temperature~\cite{Griffiths}, as a result of the dipole-dipole interaction, which, albeit weak, frustrates the tendency to ferromagnetism inherent to the exchange interaction~\cite{Arrott,Houches}. 
Thus, the dipole-dipole interaction leads necessarily to a situation of 
``avoided critical point''~\cite{Kivelson}, where the phase transition proceeds via a yet not fully elucidated and probably not universal mechanism~\cite{Kivelson,Bra,Abanov,Barci_PRB_2009,Cannas_PRB_2004,Schmalian-Wolynes_PRL2000,Note}.
At present, both the theoretical and experimental behavior of the various thermodynamic quantities in the vicinity of such an avoided critical point are uncertain. Here we demonstrate, on one side, the validity of conventional notions like critical exponents and scaling hypothesis\cite{Landau} in the vicinity of such an avoided critical point, implying the existence of a correlation length governing the phase transition. On the other side, we also establish that, when the putative critical point is {\it approached}, the non-analytic behavior transforms into an analytic one. Simultaneously, we observe that the paramagnetic-to-ferromagnetic phase transition proceeds within a background of spatially nonuniform magnetization. This observation implies that the typical period of modulation -- established by the competition between exchange and dipole-dipole interaction and characterizing the phase with nonuniform magnetization -- persists below as well as above the putative critical temperature, alongside the correlation length.
\section{Strategy} 
\vspace{-0.4cm}  
The paramagnetic-to-ferromagnetic phase transition investigated in this work occurs in  
ultrathin Fe films grown at room temperature by  molecular-beam epitaxy onto the (001) 
surface of a Cu single crystal (see Ref.~\onlinecite{Saratz_PhD_2009} for details). The system 
has macroscopic lengths within a plane but finite thickness perpendicular to the plane (typically between 1.6 and 2.0 atomic monolayers (ML)). It is magnetized perpendicularly to the plane, so that it has the right symmetry to  realize the famous Onsager critical point~\cite{OnsagerYang} of the two-dimensional (2D) Ising model. However, as shown in 
Ref.~\onlinecite{Biskup}, it also suffers the very same 
frustration foreseen for finite three-dimensional bodies so that the paramagnetic-to-ferromagnetic phase transition is associated with the appearance of stripes and/or bubble domains of opposite magnetization rather than with the formation of a spatially uniform spontaneous magnetization~\cite
{Abanov,Seul_Science_1995,Allenspach_PRL_1992,Saratz_PRL_2010,Saratz_PRB_2010,Qiu,
Back_Nature_13,Cannas_PRB_2011,Debell,Diaz-Mendez_PRB_2010,Pighin_PRB_2007}. Accordingly, the transition falls into the category of transitions with ``avoided critical point''.\\
We have followed three strategies: On one side an experimental one, consisting in performing magneto-optical Kerr effect (MOKE)~\cite{Moke} measurements of the macroscopic, spatially averaged magnetization as a function of temperature $T$ and external magnetic field $H$ applied perpendicularly to the film. Technically speaking, MOKE measures a signal which is {\it proportional} to the average magnetization~\cite{Moke}. Here we use the character ``$M$'' to identify the measured quantity, which is accordingly given in arbitrary units. In a second step, the measurement of macroscopic thermodynamic quantities was complemented with spatially resolved magnetic imaging -- performed with SEMPA\cite{Saratz_PRB_2010} --  aimed at measuring the evolution of the {\it local} magnetization during the phase transition, with submicrometer spatial resolution.
The complementary use of MOKE and SEMPA microscopy to define the phase diagram of the Fe films investigated in this work is pictorially sketched in Fig.1. 
Finally, we explored the possibility of realizing the scaling hypothesis 
in Monte-Carlo simulations of a 2D Ising model, with spins on a lattice and interacting via the familiar exchange interaction -- which promotes ferromagnetic order -- and dipolar interaction. The latter, albeit weak, is long ranged and in films magnetized perpendicularly to the plane frustrates the tendency to ferromagnetism. In the presence of the dipolar interaction, it is not a priori clear whether any scaling behavior -- explicit in the 2D Ising model with pure exchange interaction -- is left behind, nor in which region of the $(T,H)$ parameter space such scaling might be found. 
\section{Results}
\begin{figure}
\includegraphics*[width=8.5cm,angle=0]{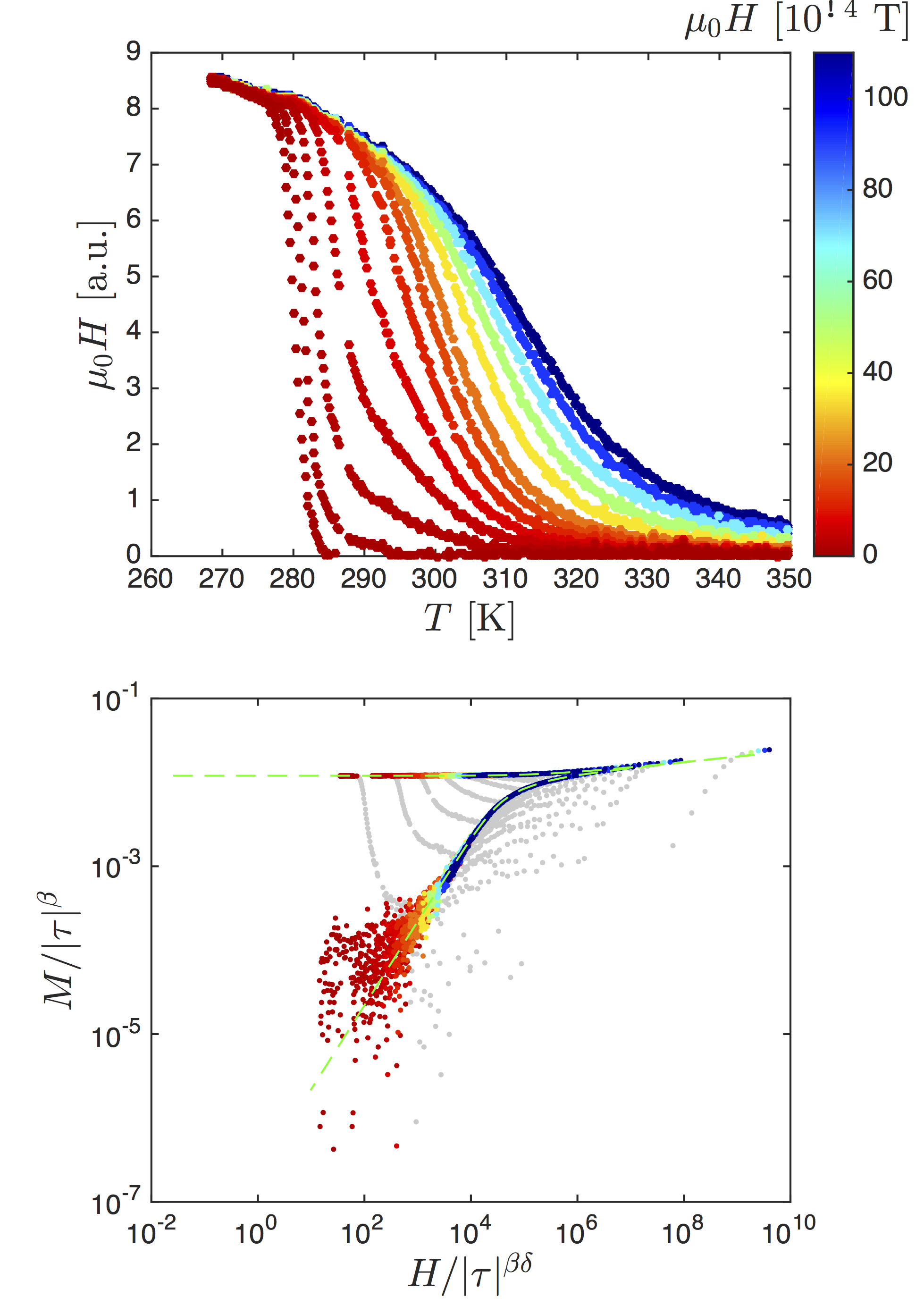}
\caption{
Top: family of experimental $M(T)$-isochamps (about $10^5$ experimental data points). The film thickness is about $1.75$ ML. The color code used for the magnetic field $H$ is given along the vertical bar. Up to $T^*$ static domains of opposite perpendicular magnetization 
are observed. Bottom: Scaling plot $M/|\tau|^{\beta}$ versus $H/|\tau|^{\beta\delta}$. The dashed green line represents the numerical scaling function, taken from Ref.~\onlinecite{Gaunt_JPhysC_1970}.
}
\end{figure}
\subsection{Scaling plots} 
Magnetization curves $M(T,H)$ are plotted as a $M(T)$-family of isochamps in Fig.2. The magnetic field was swept with a frequency varying between $10^{-1}$ and $1$ Hz. $M(H)$-curves measured at fixed temperature within this range of frequencies coincide: we assume that we are observing properties related to thermodynamic equilibrium. At higher frequencies we have detected a substantial dynamical component and the results of these studies will be reported in a separate paper. A color code, indicated along the vertical bar, is used for the values of $H$. We distinguish two extreme sets of $M(T)$-curves. On the right-hand side those corresponding to larger values of $H$, taken in a state of uniform magnetization. They show the familiar behavior of $M(T)$-curves separating out with increasing magnetic field and are suggestive of a conventional underlying critical point becoming more and more ``avoided'' with increasing applied magnetic field (this is an example of trivially ``avoided critical point''~\cite{Kivelson}). On the left-hand side the family of curves corresponding to low $H$: while the temperature is increased the average magnetization abruptly drops to almost zero as the system enters the state of static modulated order, observed directly with SEMPA (not reported here). Those images indicate that the magnetization within the domains is still large and therefore the vanishing of $M$ (\textit{global} magnetization) below $\simeq$ 300 K for small applied fields is entirely due to the cancellation of finite opposite values of the local magnetization within the domains. We point out, however, that our distinction between ``high'' and ``low'' magnetic field curves is based on our imaging of the spatial distribution of the magnetization: the macroscopic thermodynamic quantity $M(T,H)$ itself does not contain, at first glance, any specific information which could be used to classify the set of $M(T)$-curves.\\
The scaling hypothesis~\cite{Landau} states that the equation of state $M=M(T,H)$ can be simplified to a one variable relation $\tilde M={\tilde M}(\tilde \tau)$, provided the rescaled variables 
$\tilde M:= M/H^{1/\delta}$ and $\tilde \tau:= \tau /H^{\delta\beta}$ are used, $\tau\!=\!(T-T_C)/T_C$ being the so called reduced temperature. In conventional ferromagnets $T_C$ is the critical temperature at which, e.g., the \textit{critical isochamp} (order parameter) $M(T,H=0)$ vanishes. The critical exponent $\beta$ is defined via the asymptotic behavior of the critical isochamp near $T_C$, i.e. $M(T,H=0)\sim (-\tau)^\beta$.  
In our system this curve does not exist because domains form spontaneously as soon as some sizable magnetization develops locally, leading to cancellation of the  
\textit{global} magnetization. As a consequence,  $T_C$ cannot be located in a standard way, nor can $\beta$  be determined directly.  
The critical exponent $\delta$ is defined via the asymptotic behavior of the \textit{critical isotherm} $M(T_C,H)\sim H^{1/\delta}$, which -- streactly speaking -- 
does not exist either. 
However, the conventional lore of scaling (Ref.~\onlinecite{Landau}, p.485) in the critical or near critical region indicates practical ways to extrapolate values for $T_C$, $\beta$ and $\delta$ using experimental $M$-data originating within the high-temperature, non-zero field region of the parameter space, as discussed in Appendices~\ref{App_T_C} and~\ref{App_crit_exp}. 
\subsection{Phase diagram} 
\begin{figure}
\includegraphics*[width=8cm,angle=0]{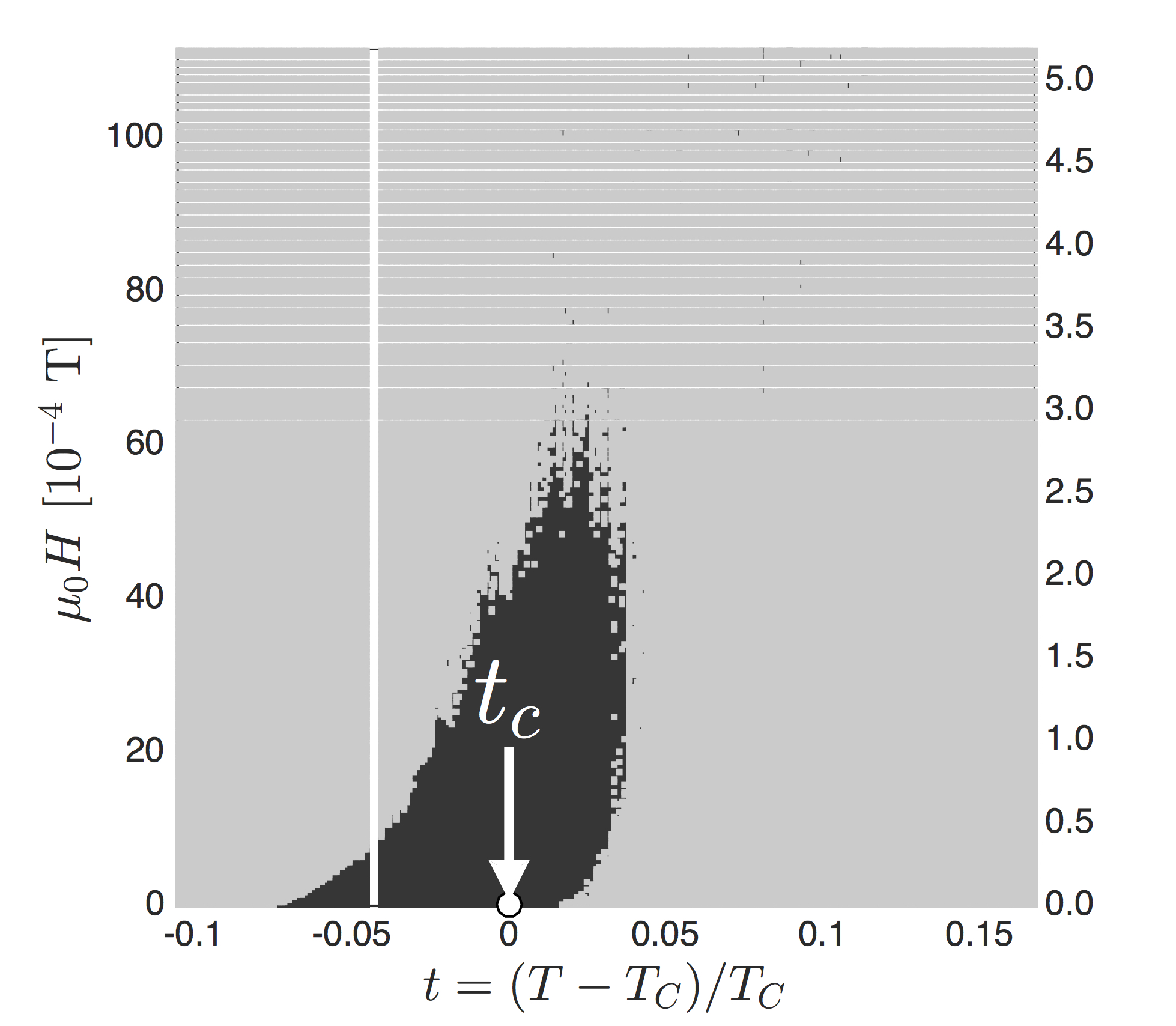}
\caption{The same experimental data points are transferred within
the $(T,H)$-parameter space. $t_c$ corresponds to the putative critical temperature used for building the variable $\tau$. The scale on the vertical right hand side gives the values of the magnetic field $\mu_0\cdot H$ in units of the saturation magnetization of Fe, $\mu_0\cdot {\cal M}_S= 2.16$ T. }
\end{figure}
Using the values of  $T_C$, $\beta$ and $\delta$ determined experimentally, the data of $M(T,H)$ can be represented in a scaling plot $M/|\tau|^{\beta}$ versus $H/|\tau|^{\beta\delta}$: The data points that collapse onto the same master curve obey the equation of state of a conventional ferromagnet in the vicinity of $T_C$.    
Therefore, ``collapsed'' and ``non-collapsed'' points can be reported in the proper location of the $(T,H)$ plane to define the phase diagram. 
In the scaling plot of Fig.2 (bottom), locations with higher density of data points can be recognized by inspection. Using an ad-hoc software, we marked the low-density points -- which we consider as ``non-collapsed'' -- in gray, while the high-density points -- which we consider as ``collapsed'' -- were left in their original color (vertical bar in Fig.2).
The gray (non-collapsed) data  and the black (collapsed) ones are subsequently transferred into their place in the $(T,H)$ plane, Fig.3, where they appear inside, respectively outside, a bell-shaped region marked in gray. To represent the phase diagram in Fig.3 the  variable $t:=(T-T_C)/T_C$ is used. Notice that many reddish color-coded points appear in the lower left-hand section of the scaling plot Fig.2 (bottom). They correspond to high temperatures and low applied magnetic fields and they are scattered because of noise, resulting in a small density. Despite this, we consider them as ``collapsed'', because they are spread along the continuation of the scaling function of the 2D Ising model (the dashed line in Fig.2 (bottom)). The bell-shaped region starts on the left with the boundary 
line marking the transition from (static) modulated-to-uniform phase, recorded by direct SEMPA imaging of the spatial distribution of the magnetization. It continues beyond  the range of temperature and magnetic fields where a static modulation of order is observed and {\it even beyond} $t=0$, that is above $T_C$. Evidently, the static magnetic domains~\cite{Saratz_PRL_2010} imaged with SEMPA represent only a small portion of the ferromagnetic transition.

Remarkably, the singular behavior of $M=M(T,H)$ expected for a conventional ferromagnet while $T_C$ is approached gives way to an analytic behavior within the ``gray  zone''.  In particular, the $M$ versus $H$ dependence for low magnetic fields is perfectly linear within the ``gray'' zone, i.e. at sufficiently low magnetic fields a singular behavior, characterized by some non-linear $M$ versus $H$ relationship, does not develop. It appears, therefore, that the non-analytic behavior is precisely suppressed when the putative critical point is approached.  \\

\subsection{Griffiths-Widom representation} 
\begin{figure}[b!]
\includegraphics*[width=8cm,angle=0]{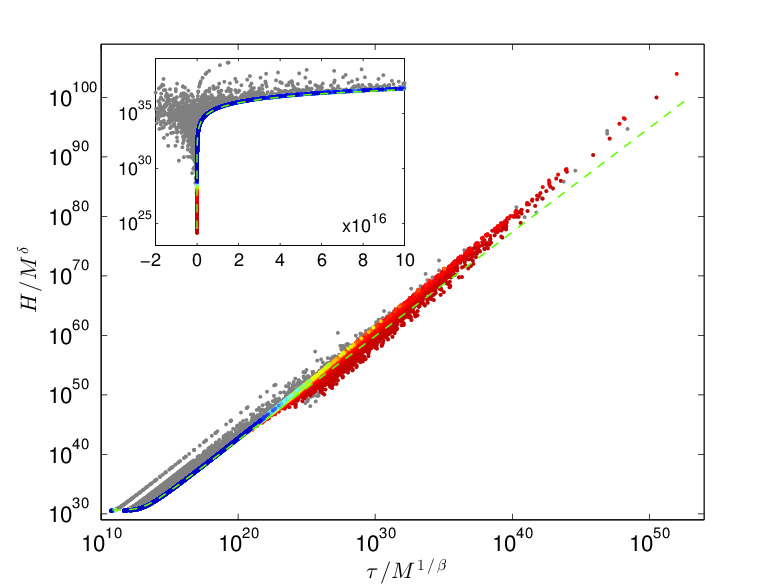}
\caption{\label{SM_fig5}  Scaling plot in the Griffiths-Widom representation $H/M^\delta$ vs $x=\tau/ M^{1/\beta}$. Data and color codes are the same as in Fig.2. 
The scale is log-log in the main panel and log-linear in the inset. The green dashed line represents the theoretical scaling function $h(x)$ given in Ref.~\onlinecite{Gaunt_JPhysC_1970} for the 2D Ising model (shifted by non-universal, constant scaling factors); note that collapsed data points overlap to this line throughout the range $-6\cdot 10^{12}\lesssim x \lesssim 10^{23}$. The slight deviations for $x>10^{40}$ are due to the experimental value of the exponent $\gamma$ being slightly different from the one expected for the 2D Ising
(remember that $\gamma$ is the slope of the graph of Fig.4 for very large values of $x$).}
\end{figure}
An alternative way with respect to Fig.2~(bottom) to express the equation of state of a ferromagnet is the Griffiths-Widom representation~\cite{Griffiths-Widom}: 
\begin{equation}
\label{scaling_fct}
\frac{H}{M^\delta} = h\left(x\right)
\end{equation}
with $x=\tau /M^{1/\beta}$. In Fig.4 the same data points shown in Fig.2~(bottom) are plotted in this representation with the same color coding.
Particularly clear is how gray points deviate from scaling for the data plotted in the inset ($-2\cdot 10^{16}\leq x \leq 10^{17}$). In the main frame
the non-collapsed, gray points decorate the colored \textit{line} of clearly collapsed points up to $x\sim 10^{23}$. For larger values of $x$, the main line broadens as well: these are the high temperature, low-field ``noisy'' data points already discussed in relation to  Fig.2. In the Griffiths-Widom representation data collapsing is realized over {\bf forty} orders of magnitude with respect to the $x$-variable and {\bf eighty} orders of magnitude with respect to the variable $H/M^\delta$. The theoretical scaling function $h(x)$ given in Ref.~\onlinecite{Gaunt_JPhysC_1970} for the unfrustrated 2D Ising model is plotted as a green dashed line. Colored, collapsed points follow this curve very well up to  $x\lesssim 10^{23}$ (note the remarkable agreement in the inset).

\subsection{Monte-Carlo simulations}
\begin{figure}
\includegraphics*[width=8cm,angle=0]{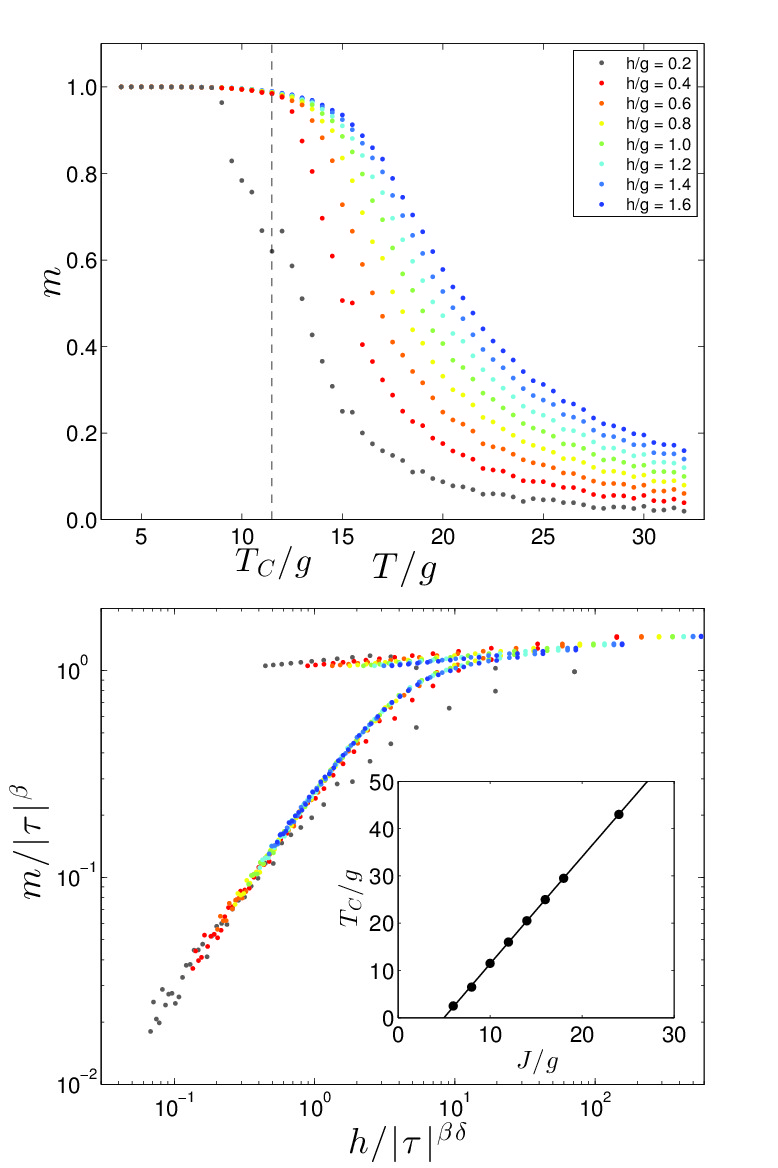}
\caption{
Monte-Carlo simulations. A family of $m(T)$-isochamps (top), computed for $J/g=10$, $L_x=L_y=120$ and different values of the magnetic field $h$, collapses onto one single scaling function (bottom) when the magnetic field exceeds the critical value for entering a spatially uniform state of magnetization. The non-collapsed data points correspond to a magnetic field below the threshold. Some simulations taking $L_x=L_y=200$ were also run to check for possible finite-size effects. The inset at the bottom shows the characteristic temperature $T_C$, used for optimize data collapsing in the Monte-Carlo scaling plots, as a function of $J/g$.  
}
\end{figure}

We consider the Hamiltonian~\cite{Pighin_PRB_2007}
\begin{equation}
{\cal H}= - J \sum_{\langle i,j \rangle} S_i S_j + g\, \sum_{(i,j)} \frac{S_i
S_j}{r^3_{ij}} + h \, \sum_i S_i \label{Hamilton1}
\end{equation}
\noindent where the spins $S_i=\pm 1$  are defined on a square lattice with
$N = L_x \times L_y$ sites and periodic boundary conditions (Ewald sums technique was used to handle them). The first term in Eq.~\eqref{Hamilton1} represents the short-range exchange interaction (the sum runs over all pairs of nearest-neighboring sites). The second term represents the long-range dipolar interaction in the Ising limit, the corresponding sum running over {\it all} pairs of distinct sites on the lattice. The third term represents the Zeeman energy. The Hamiltonian in Eq.~\eqref{Hamilton1} is far from describing ultrathin Fe films on Cu(100) in detail. However, it is considered to build a model Hamiltonian capturing realistically competing interactions and the avoided critical point. Accordingly, it has been found to display a variety of modulated phases~\cite{Debell,Diaz-Mendez_PRB_2010,Pighin_PRB_2007,Cannas_PRB_2006} within some region of the $(T,h)$ parameter space, even for those moderate values of the ratio $J/g$ where Monte-Carlo simulations are practicable~\cite{footnote1}. In this particular study we have explored the phase diagram outside this region, with the aim of searching for some residual scaling behavior. In a first step we have determined a limiting field $h_c$ above which the system is in a spatially uniform state, within the temperature range considered by the simulation. $h_c$ was determined from the behavior of the average spin polarization $m:= (\sum_i \langle S_i\rangle)/N$ as a function of temperature
for fixed values of $h$, according to a zero-field-cooled--field-cooled (ZFC-FC) protocol
\cite{footnote2}. In a second step we have computed isochamps during the heating part of 
the ZFC-FC cycle, shown on the top of Fig.5 for $J/g=10$ and $h_c/g=0.32 \pm 0.02$. The 
spreading of the isochamps with different values of the magnetic field is similar to the 
experimental one reported in the top of Fig.2. In a third step we attempted scaling plots, 
trying to bring the isochamps to collapse onto one single curve. It turned out that 
collapsing could be indeed realized for $h>h_c$ (colored in the bottom of Fig.5) and a 
significant departure from collapsing was observed for $h<h_c$ (gray points 
corresponding to $h/g=0.2$), i.e. inside the region of the phase diagram where nonuniform 
magnetization appears. For building the scaling plots of Fig.5 we used the known 2D Ising 
critical exponents, but the value for the putative critical temperature $T_C$, required to 
build the reduced temperature $\tau$, cannot be read out in an obvious way from the set of 
isochamps of Fig.5. We found, for a practicable set of $J/g$ values, that the best 
data collapse could be realized assuming a transition temperature given in the inset of 
Fig.5 (bottom) as a function of $J/g$. Remarkable about this temperature 
is that it can be expressed as the transition temperature for the ``pure'' 2D Ising model 
subtracted by a value which corresponds to about $11.2\, g$ (see Inset of Fig.5 (bottom)). We recall that, within a simple mean-field approach, the presence of the dipolar interaction in a uniformly magnetized state 
reduces the transition temperature by an amount corresponding to about $4\pi g$! 
The picture emerging from Fig.5 is in line with the experimental outcome and underlines the 
realization of scaling and power laws in a situation where the critical point is avoided.  
\section{Discussion}
\begin{figure}[b!]
\begin{center}
\includegraphics*[width=8.cm,angle=0]{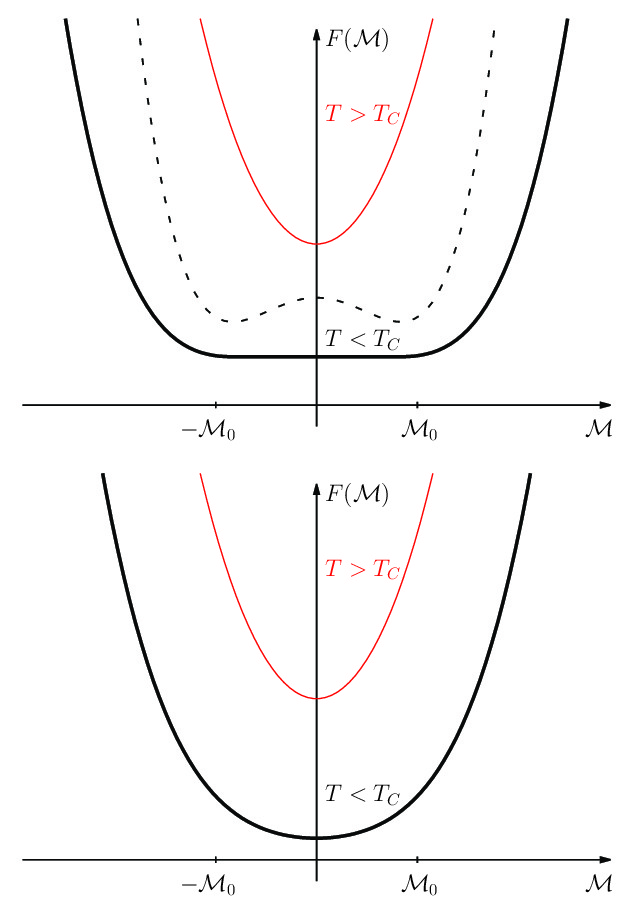}
\caption{
Top: The Free energy density $F({\cal M})$ for a ferromagnet at $T$ larger than the Curie temperature $T_C$ (thin solid line) and $T$ below $T_C$ (thick solid line). Notice that  the flat portion is slightly modified into two flat minima at $\pm M_0(T)$ if the three dimensional body is large but finite (dashed line)\protect\cite{Griffiths2}. Bottom: in the presence of the dipolar interaction, the flat portion (the two minima) changes into one single minimum at ${\cal M}=0$.}
\end{center}
\end{figure}
The familiar description of a second-order phase transition foresees that at 
sufficiently large temperatures $T$ the free-energy density $F({\cal M})$ in the variable ${\cal M}$ (average magnetization) has a minimum at ${\cal M}\!=\!0$ (Fig.6, top). When the temperature is lowered below a critical value $T_C$, the graph of $F({\cal M})$ acquires a flat portion (Fig.6, top) between two limiting values $\pm {\cal M}_0(T)$ defining a situation of spontaneously broken symmetry with spatially uniform magnetization $\pm {\cal M}_0(T)$. The present paper is in contrast to this picture because it finds that the flat portion is replaced by a minimum of $F({\cal M})$ at ${\cal M}\!=\!0$ for {\it any} temperature (Fig.6, bottom).
Notice that we are not the first to point out the difficulty of translating the familiar description of second-order phase transitions~\cite{Landau} to the paramagnetic-to-ferromagnetic transition~\cite
{Griffiths,Arrott}. But the present results pinpoint some essential elements of the ``gray zone'' replacing the critical point which were not recognized explicitly yet: 
$i$) the recovery of the familiar properties of scaling and power laws, albeit {\it outside} the 
gray zone and $ii$) the persistence {\it above} 
the putative critical temperature of the phase of the magnetic-domain patterns, accessible to spin-sensitive scanning probe techniques {\it only} below $T_C$ and when 
domains are static (on the time scale of a specific scanning-probe experiment). \\
It is important to mention at this point the vast literature~
\cite{Poki,Amn} describing the cross-over from the exchange controlled criticality to the 
dipolar-controlled criticality within a {\it very narrow} temperature range in the 
vicinity of the critical point. Notice that the cross-over considered in this literature {\it preserves} the critical point (albeit with slightly modified critical exponents), in contrast to our observations. However, the cross-over scenario refers to a situation of uniform spontaneous magnetization which is actually forbidden for finite three-dimensional bodies. We can, in principle, envisage special  
situations where a uniform magnetization is possible, such as the infinite-volume limit underlying the cross-over scenarios~\cite{Poki,Amn}, but also an inefficient domain nucleation that keeps the body in a metastable state of uniform magnetization, or some very particular shapes (not covered by Ref.~\onlinecite{Griffiths}) which energetically penalize the formation of domains. However, the true absence of any domain, both below as well as {\it above} the putative critical point, must be verified specifically\cite{AA1}, as these situations must be regarded as exceptional while the rule is rather a ``gray zone'', i.e. an avoided critical point. Unfortunately, it is very difficult to establish, on the base of experimental and theoretical data related to macroscopic thermodynamic quantities alone, the extent of the gray zone~\cite{AA2}.\\
In summary, the recovering of the scaling hypothesis outside the ``gray region'' means, on 
one side, that scaling properties and critical exponents, referring to macroscopic 
thermodynamic quantities, are not necessarily a proof of the existence of a critical 
point. On the other side, their emergence in a situation of avoided criticality, where the 
phase transition actually might even be a purely dynamical one~\cite
{Note}, sets some well-defined boundaries on future realistic 
models of the ferromagnetic phase transition (and of second-order phase transitions in 
general). These models will have to take into account the presence of (at least) two 
mesoscopic spatial scales: the correlation length, whose divergence at the putative 
critical point is an essential ingredient for the current understanding of critical 
phenomena, {\it and} the dipolar-induced period of modulation characterizing the phase with domain patterns.   \\

\textit{Acknowledgments} -- 
We thank Thomas B\"ahler for technical assistance, G. M. Graf, A. Giuliani, O. V. Billoni and S. Ruffo for helpful discussions as well as the Swiss National Science Foundation, ETH Zurich, and CONICET (Argentina) for financial support.

\appendix
\section{Experimental determination of $T_C$\label{App_T_C}}
%
\begin{figure}[b]
\label{S1}
\includegraphics*[width=8cm,angle=0]{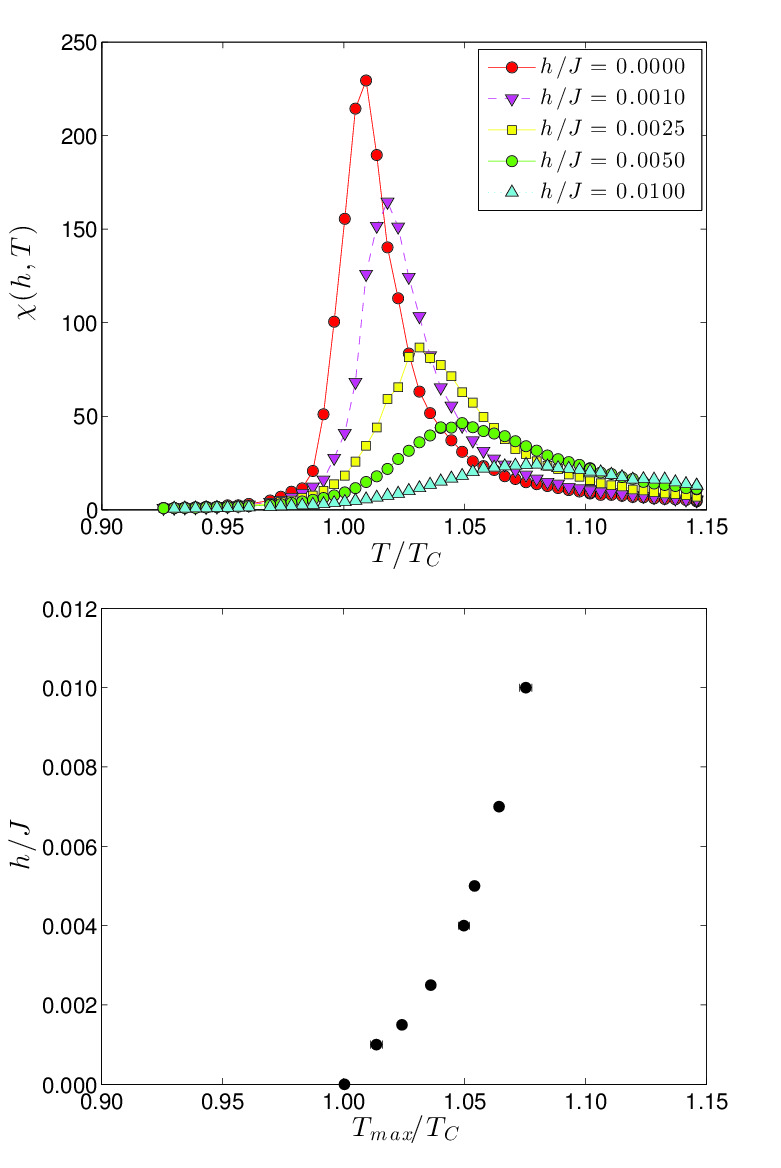}
\caption{Monte-Carlo simulations for the pure 2D Ising model.  
Left: Plot of the magnetic susceptibility $\chi(T,h)=\partial m/\partial h$ as a function of $T/T_C$ for a lattice of linear size  $L_x=L_y=120$; we recall that $T_C=2.269 \,J$, assuming $k_B=1$ \cite{OnsagerYang}; the selected values of $h/J$ are given in the figure legend. 
Right: $h/J$ versus $T_{max}(h)/T_C$; 
for each field, the value of $T_{max}(h)$ was extrapolated with finite-size scaling using lattices of size $L_x=L_y=$48, 64, 96, 128 and 200.}
\end{figure}
\begin{figure}[t]
\includegraphics*[width=8cm,angle=0]{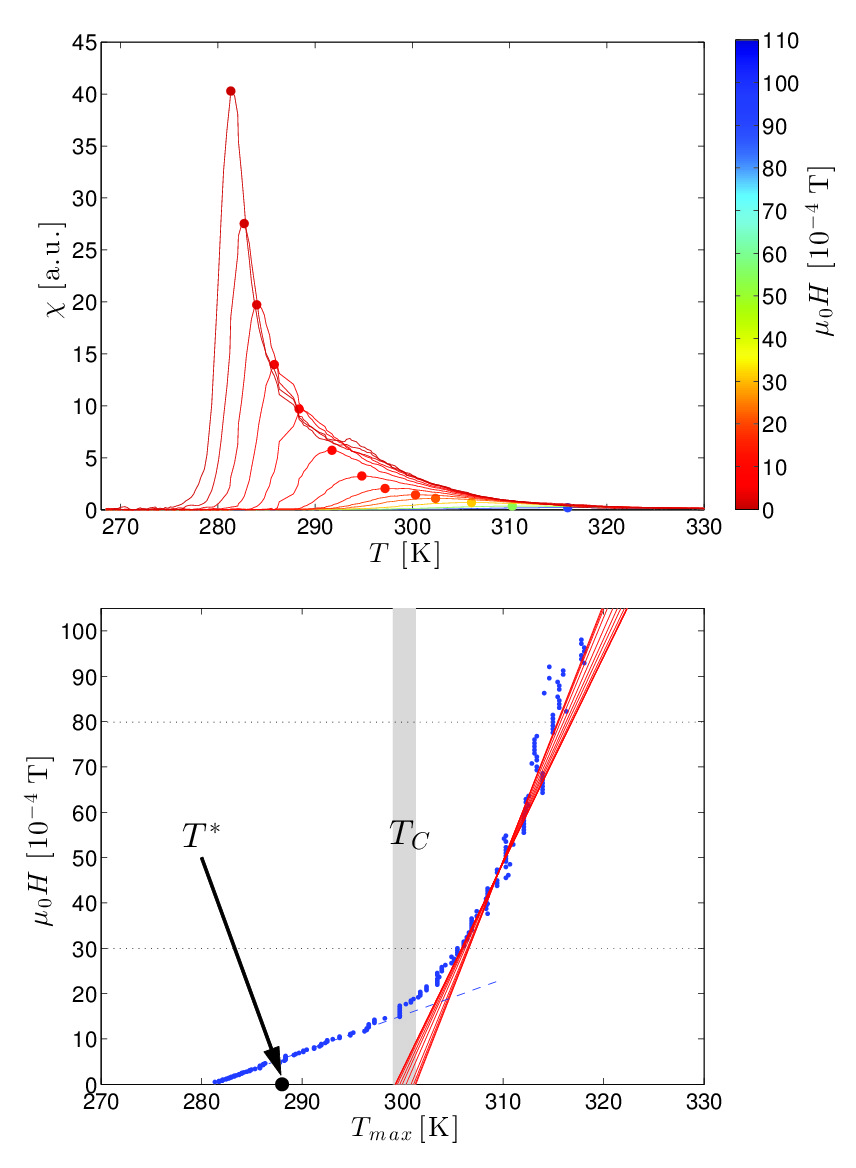}
\caption{
\label{SM_fig2} 
Top: Plot of the magnetic susceptibility $\chi(T,H)=\partial M(T,H)/\partial H$ of Fe/Cu(001) films as a function of temperature obtained as described in the text. Different colors correspond to different applied fields (see color map on the right). 
Bottom: Plot of $H$ vs $T_{max}(H)$ obtained from the experimental susceptibility as described in the text; the blue dashed line is a guide to the eye for the maxima marking the re-entrant transition
from the patterned to the uniform phase; linear fittings used to determine $T_C$ are drawn as red lines; horizontal dotted lines indicate the region where fittings were performed; the 
shadowed vertical stripe corresponds to the estimate of $T_C=300 \pm 1$ K; the temperature $T^*$ above which SEMPA images become contrastless is also indicated.  
}
\end{figure}
One useful property of a conventional second-order phase transition is that the plot of the magnetic susceptibility  $\chi(T,H)=\partial M(T,H)/\partial H$ 
as a function of the temperature has a maximum at a temperature $T_{max}(H)$ that approaches  $T_C$ as $H$ approaches zero~\cite{Change-Lee}. Figure 7 (top) shows the susceptibility $\chi(T,H)$ of the ``pure'' (i.e. without dipolar interactions) 2D Ising model computed by means of Monte-Carlo simulations. For any $H\not=0$ the maximum value of $\chi(T,H)$ is finite but -- as expected -- it is higher the weaker the fields are. 
Notice that the true approach of $T_{max}(H)$  to $T_C$ may not necessarily be linear in $H$, as pointed out in Ref.~\onlinecite{Change-Lee}, but our numerical simulations (bottom of  Fig.7) show that a linear extrapolation of $T_{max}$ toward $H=0$ gives a fairly accurate estimate of $T_C$.\\
\noindent From the experimental data we obtained the $\chi(T,H)$ by first treating the raw $M(T,H)$ data with a Savitzky-Golay finite-impulse-response smoothing filter implemented in MATLAB. After filtering, the derivative $\chi(T,H)=\partial M(T,H)/\partial H$ 
could readily be obtained. In Fig.8 (top) some of the resulting susceptibility curves are shown for selected values of $H$. The experimental $H(T_{max})$ is plotted at the bottom of Fig.8. This graph, in contrast to the one of the ``pure'' 2D Ising model, shows two distinct regimes, depending on the range of 
$H$. We discuss first the low temperature regime. When cooling in weak enough fields, the system first enters the ``gray zone'', (see the phase diagram in Fig.2 of the main document), without displaying any anomaly in the susceptibility in correspondence of this transition. Upon further cooling a re-entrant transition from the patterned to the uniform phase is encountered~\cite{Saratz_PRL_2010,Saratz_PRB_2010,port}: this second transition is accompanied by a sharp maximum in the susceptibility (abrupt increase of the magnetization). This type of maxima -- highlighted by a blue dashed line in Fig.8 (bottom) -- when extrapolated to $H=0$ leads to a temperature at which the sample consists of very large stripes carrying opposite but almost saturated values of the magnetization~\cite{Saratz_PRL_2010,Saratz_PRB_2010,port}. Above this temperature, the sample keeps the modulated order up $T^*$, where domains become mobile but the magnetization within them is still substantial. Accordingly, $T_C$ must be above $T^*$. On the other side, for larger fields (the right-hand side portion of the graph) the ``gray zone'' is never entered and the system is in a uniform state:  $H(T_{max})$ for larger $H$ was therefore taken to extrapolate towards a putative $T_C$. 
Several linear fittings were performed by choosing different ranges of the field $\mu_0\cdot H$ between $30\cdot 10^{-4}$T and $80\cdot 10^{-4}$ T. The different fittings produced the family of red lines (bottom of Fig.8) from which the error on the $T_C$ was estimated. The experimental $T_C$, for this particular sample, is $300 \pm 1$ K. Notice that the extrapolated $T_C$ lies within the cross-over range (K) $ 298\!\lesssim \! T \!\lesssim \!302$, in which the ``blue-line'' types of maxima transform into the ``red-line'' types of maxima.  
\section{Experimental determination of $\beta$ and $\delta$\label{App_crit_exp}}
The critical exponent $\beta$ can be deduced from the experimentally determined values of the exponents $\gamma$ and $\delta$, using the relation $\beta\,\delta\!=\!\beta \!+\!\gamma$. The exponent $\gamma$ determines the magnetization in the region of weak fields (Eq. 148.8. in Ref.~\onlinecite{Landau}) according to
\begin{equation}
M\sim \frac{H}{\mid \tau\mid ^{\gamma}}\,.
\end{equation}
The exponent $\delta$ determines the magnetization in the region of strong fields according to the relation
\begin{equation}
M\sim H^{1/\delta}
\end{equation}
(see Eq. 148.10 in Ref.~\onlinecite{Landau}). The notion of ``weak'' and ``strong'' field is, of course, dependent on which temperature interval is addressed.\\
%
\begin{figure}[t!]
\includegraphics*[width=8cm,angle=0]{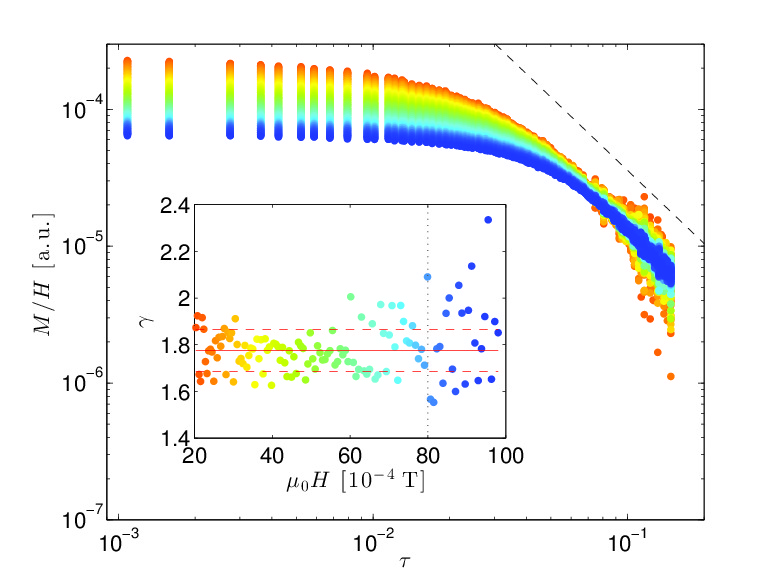}
\caption{
\label{SM_fig3} Log-log plot of the ratio  $M/H$ vs $\tau=(T-T_C)/T_C$ (with $T_C=300$ K); the dashed black line is a guide to the eye with slope equal to the mean value of $\gamma$. Different isochamps are plotted with different colors; specific values of the field $H$ can be identified from the horizontal axis of the inset, where fitted values of $\gamma$ obtained for different fields are plotted (see the text); the solid and dashed horizontal lines indicate the mean value of $\gamma$ and the standard deviation from this average, respectively; the vertical dotted line marks the largest field $\mu_0\cdot H=80\cdot 10^{-4}$ T used to determine $\gamma$.}
\end{figure}
\begin{figure}[b!]
\includegraphics*[width=8cm,angle=0]{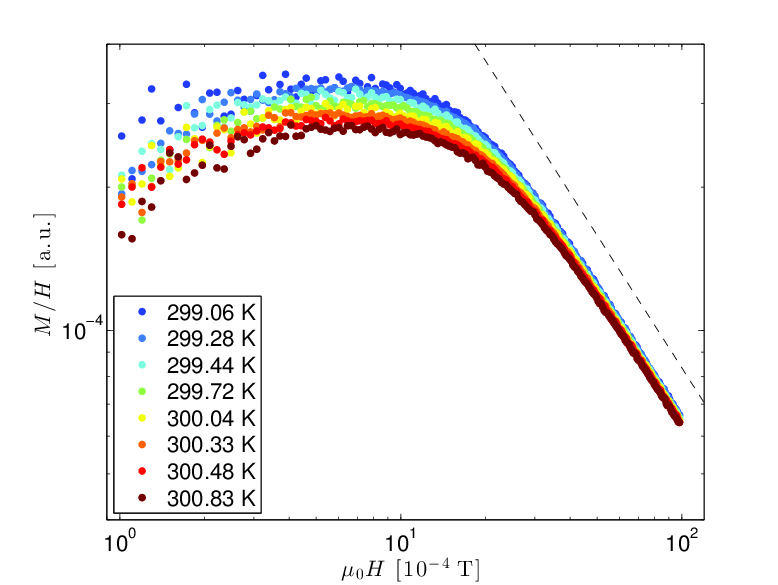}
\caption{Log-log plot of the ratio  $M/H$ vs $H$. Different isotherms are plotted with different colors (specific values of $T$ are given in the legend). 
The dashed black line is a guide to the eye with slope equal to mean value of $(1-\delta)/\delta$ (fitted exponents).} 
\end{figure}
In Fig.9 we plot $\log(M/H)$ versus $\log \mid \tau\mid $ for different values of $H$, with the color code indicating the values of $H$ (see horizontal scale in the inset). At sufficiently high temperatures, a region of the graph emerges where all curves for different magnetic fields almost collapse onto one single straight line, and thus fulfill the scaling properties required by Eq.~(1) for the quantity $M/H$. The negative of the slope of the resulting straight line is the sought-for exponent $\gamma$. Several linear fittings were performed for fixed fields ranging from $20\cdot 10^{-4}$ T to $80\cdot 10^{-4}$ T; these independent determinations of $\gamma$ are shown in the inset. 
After averaging, for $T_C=300$ K we obtain $\gamma=1.78\pm 0.09$, with the error given by the standard deviation of the mean. 
As consistency check, the whole procedure was repeated varying the value of $T_C$. The standard deviation of the mean values of $\gamma$ goes through a minimum in the range (K) $298\!\le\!T_C\!\le\!301$, which is, accordingly, the range where the ``best collapsing'' of the $M/H$-curves is realized. When $T_C$ is varied in this interval, $\gamma$ ranges from 1.7 to 1.9.\\
The $\log(M/H)$ versus $\log H$ plot of Fig.10 in the temperature range (K) $299\!<\!T\!<\!301$, reveals a low-field region where the curves saturate to an almost constant vale, indicating the linearity of $M$ versus $H$ for weak fields. In the strong-field region the graphs are observed to almost collapse onto a single straight line, the slope of which amounts to $(1-\delta)/\delta$, consistently with Eq.~2.
From these slopes fitted for different $T$ in the appropriate regime of Fig.10 we estimate $\delta = 13 \pm 2$.


\begin{thebibliography}{10}
\bibitem{footnote2}
After having prepared the system in a uniform saturated state (with $m=1$), we let the temperature increase from very low values up to a predefined
$T_f$, lying above the transition from the modulated to the paramagnetic phase for $h=0$.
Then, we stopped  the simulation and, starting from the final configuration, the system was cooled down to the original temperature. When $h<h_c$ a strong hysteresis in the $m(T)$ curves was observed, below the temperature where modulated phases develop. When $h>h_c$, instead, the curves were completely reversible in the whole temperature range. In this way we could estimate $h_c$ for rather large values of $J/g$ (up to $J/g=10$), without the computational cost of complete phase-diagram calculations.
\bibitem{AA1}
Our results have shown that the size of domains can change from few $100\mu$m to few $10$ nm within a narrow temperature range in the transition region, so that a state which appear to be uniform sufficiently below $T_C$ cannot be automatically assumed to remain uniform when the transition temperature is approached.
\end{thebibliography}
\end{document}